\documentclass[12pt]{article}
\pdfoutput=1

\usepackage{paper2e}
\usepackage{mydefs2e}
\usepackage{xspace}
\usepackage{graphicx}
\usepackage{amssymb}

\newcommand{\Sla}[1]%
{\kern0.12em{\raise.15ex\hbox{$/$}\kern-.74em #1}}% Feynman slash

\newcommand{\MS}{\ensuremath{M_{\rm SUSY}}\xspace}
\newcommand{\LC}{\ensuremath{\La_{\rm CTC}}\xspace}
\newcommand{\MP}{\ensuremath{M_{\rm P}}\xspace}

\newcommand{\beqa}{\begin{eqnarray}}
\newcommand{\eeqa}{\end{eqnarray}}
\newcommand{\lp}{\left(}
\newcommand{\rp}{\right)}

\begin{document}

% =============================================================================
% Title page
% =============================================================================
\begin{titlepage}
%\preprint{}

\title{Flavor in Minimal %\\\medskip
Conformal Technicolor}

\author{Jamison Galloway$^{*\dagger}$,\ \ 
Jared A. Evans$^\dagger$,\\
Markus A. Luty$^\dagger$,\ \ 
%and\ \ 
Ruggero Altair Tacchi$^\dagger$}

\address{$^*$Dipartimento di Fisica, Universit\`a di Roma ``La Sapienza"\\
and INFN Sezione di Roma, I-00185 Roma}

\address{$^\dagger$Physics Department, University of California Davis\\
%One Shields Avenue\\
Davis, California 95616}

\begin{abstract}
We construct a complete, realistic, and natural UV completion
of minimal conformal technicolor that explains the origin of
quark and lepton masses and mixing angles.
As in ``bosonic technicolor,''
we embed conformal technicolor in a supersymmetric theory,
with supersymmetry broken at a high scale.
The exchange of heavy scalar doublets generates higher-dimension
interactions between technifermions and quarks and leptons
that give rise to quark and lepton masses at the TeV scale.
Obtaining a sufficiently large top quark mass requires
strong dynamics at the supersymmetry
breaking scale in both the top and technicolor sectors.
This is natural if the theory above the supersymmetry breaking
also has strong conformal dynamics.
We present two models in which the strong top dynamics
is realized in different ways.
In both models, constraints from flavor-changing effects can be 
easily satisfied.
The effective theory below the supersymmetry breaking scale
is minimal conformal technicolor with an additional light
technicolor gaugino.
We argue that
this light gaugino is a general consequence of
conformal technicolor embedded into a supersymmetric theory.
If the gaugino has mass below the TeV scale it will give
rise to an additional pseudo Nambu-Goldstone
boson that is observable at the LHC.
\end{abstract}

\end{titlepage}

% =============================================================================
\section{Introduction}
\label{sec:intro}
% =============================================================================
Conformal technicolor \cite{CTC} is a solution of the hierarchy problem
in which the electroweak scale is determined by the soft breaking of
conformal invariance \cite{CTClatticeidea}.
The minimal theory of this kind is $SU(2)$ technicolor with
additional technifermions uncharged under the standard
model gauge group,
such that the $SU(2)$ gauge coupling has a strong
conformal fixed point.
The strong conformal symmetry is broken softly by a mass term for the 
sterile technifermions, which results in the breaking of electroweak symmetry.
We will refer to this theory as minimal conformal technicolor (MCTC).

This theory breaks electroweak symmetry via strong dynamics near
the TeV scale.
Two potentially serious problems with such scenarios are
precision electroweak constraints and flavor physics.
The precision electroweak constraints in MCTC were
discussed in \Ref{CTCmin}.
This model has a space of vacua given by
$SU(4)/Sp(4) \simeq SO(6)/SO(5)$,
which contains a pseudo Nambu Goldstone Higgs.
The theory can have a standard model-like fit to the precision
electroweak data with a light composite Higgs at the price
of $\sim 10\%$ fine tuning.
Alternatively, the theory may live in the technicolor vacuum
provided the $S$ parameter is less than about half the QCD
value and there is a positive contribution to the $T$
parameter.

The subject of this paper is a detailed investigation of flavor in
MCTC.
Electroweak symmetry breaking is communicated to the fermions
via higher dimension interactions such as
\beq[topop]
\De\scr{L}_{\rm eff} \sim \frac{1}{\La_t^{d-1}}
(q_3 \tilde{t}) \scr{H} + \hc
\eeq
where $q_3$ and $\tilde{t}$ are the third generation quark
fields, and
$d$ is the UV dimension of the operator
$\scr{H} = \psi \tilde\psi$ that breaks electroweak symmetry.
In conformal technicolor (as in ``walking'' technicolor \cite{walking})
$d < 3$, allowing flavor to be generated at a higher scale.
We focus on conformal rather than walking dynamics because
conformal dynamics is more robust and plausible.%
\footnote{Walking dynamics is plausible in large-$N_c$ theories
at the end of the conformal window \cite{largeNwalk},
but in such theories
the $S$ parameter is proportional to $N_c$, and is generically
too large.}

Conformal dynamics of non-abelian gauge theories is now being
actively investigated on the lattice \cite{conformallattice}.
In particular, the dimension $d$ can be readily determined in
lattice simulations \cite{CTClatticeidea}
and measurements are starting to appear
\cite{CTClatticemeasure}.
Interest in conformal theories in the lattice community is
growing and we expect significant progress in the near future.
One can also obtain bounds on $d$ using rigorous results of
conformal field theory \cite{CFTbounds}.
The idea is that if
$d$ is too close to $1$, then the operator $|\scr{H}|^2$
has dimension close to 2, and the fixed point is IR unstable.
One can use this to obtain rigorous bounds on $d$,
but for theories with global symmetries (the case of interest)
the bounds are very weak \cite{CFTboundsglobal}.

For what values of $d$ can we get a realistic theory
of flavor?
The main constraint is that the top quark mass is generated
by the interaction in \Eq{topop}.
If we choose the coefficient such that $\La_t$ is the scale
where the operator becomes strongly coupled in the UV, 
we have
\beq
m_t \sim 4 \pi v
\left( \frac{\LC}{\La_t} \right)^{d-1},
\eeq
where $\LC \sim 4\pi f$ is the $SU(4)$ breaking scale.
This gives
\beq[Lambdatestimate]
\La_t \sim 4 \pi v
\left( \frac{m_t}{4\pi v} \right)^{-1/(d-1)}
\sim \begin{cases}
10\TeV & $d = 3$ \cr
40\TeV & $d = 2$ \cr
600\TeV & $d = 1.5$ \cr
\end{cases}
\eeq 
where we (conservatively) use
$f = v$ in the numerical estimates.
(Note that $m_t$ renormalized at $4\pi v \sim 2\TeV$
in a theory with no light Higgs is $133\GeV$.)
New physics that generates the operator \Eq{topop} 
must appear at or below the scale $\La_t$.
This theory has flavor-dependent couplings, but
must not generate unacceptably large flavor-changing
effects.
To know whether this is possible for a given value of $d$
we need a more fundamental theory that generates
\Eq{topop}.

In this paper, we address this question 
by constructing natural and realistic theories of flavor
that give a UV completion of MCTC.
We recall that MCTC
is a theory of electroweak symmetry breaking with gauge group
\beq
SU(2)_{\rm CTC} \times SU(2)_L \times SU(2)_R
\eeq
where $Y = T_{3R}$.
The theory has fermions
\beq
\bal
\psi &\sim (2, 2, 1),
\\
\tilde\psi &\sim (2, 1, 2),
\eal
\eeq
plus enough additional ``sterile technifermions''
(charged under $SU(2)_{\rm CTC}$ but not $SU(2)_L \times SU(2)_R$)
so that the gauge coupling has a strongly coupled fixed point.
The existence of such a fixed point is expected at the end of
the conformal window, and is under active investigation in the
lattice gauge theory community.
Conformal symmetry is then softy broken by mass terms
for the sterile technifermions.
At the scale of the soft breaking the theory confines
and spontaneously breaks the $SU(4)$ chiral symmetry
acting on $\psi, \tilde\psi$ via a  fermion condensate.
Weak interactions that explicitly break $SU(4)$ give rise to
a potential can align the condensate to give either a light
PNGB Higgs or a technicolor-like theory below the TeV scale.

In this paper, we embed this theory into a supersymmetric theory, with
SUSY broken at a scale $\MS \gg \mbox{TeV}$,
which plays the role of the flavor scale.
It is natural for SUSY to be broken at the same scale in
the technicolor and standard model sectors
(\eg\ in gravity mediation), therefore the superpartners
of the standard model particles also have masses of order \MS.
Effective interactions like \Eq{topop} that generate the quark and
lepton masses arise from the exchange of elementary scalars 
at \MS.
These scalars have the quantum numbers of Higgs bosons, but they
do not have VEVs and their role is that of flavor messengers.
The quark and lepton masses and mixings are then proportional
to ordinary Yukawa couplings.
Because $\MS \gg \mbox{TeV}$, the flavor-changing effects from
squark and slepton mixing are suppressed or absent.
For completely arbitrary $CP$-violating squark masses, we need
$\MS \gsim 10^3\TeV$, but $\MS \gsim 10\TeV$ may be sufficient
if the soft SUSY breaking preserves $CP$ and has some degree of alignment.
SUSY and technicolor may therefore solve each others' flavor problems.
This is the basic idea of ``bosonic technicolor'' \cite{bosonicTC}.
(see also \Refs{TCHiggs}).
SUSY has also been used to UV complete little Higgs theories
in \Refs{intHiggs}, and our models have some features in common
with these models.

However, this idea by itself is not sufficient to explain the
observed value of the top quark mass.
In this scenario, we have
\beq[mtbtc]
m_t \sim 4\pi v
\left( \frac{y_t}{4\pi} \right)
\left( \frac{y_{\rm TC}}{4\pi} \right)
\left( \frac{4\pi f}{\MS} \right)^{d-1}.
\eeq
where $y_t$ and $y_{\rm TC}$ are the top and
technicolor Yukawa couplings at the scale \MS.
\Eq{mtbtc} states that the top quark mass is given by its
strong-coupling value ($\sim\mbox{TeV}$) times all suppression factors 
that make it less than strongly coupled at the scale
$4\pi f$ where chiral symmetry breaking occurs.
This estimate assumes that the operator $\psi \tilde{\psi}$
has dimension $d$ immediately below \MS.
For $d < 3$, this requires that the technicolor coupling is
at a non-supersymmetric strong fixed point right below
\MS.
If instead the technicolor coupling is weak at the the scale
\MS and runs to a strong fixed point, the top mass
will have additional suppression compared to \Eq{mtbtc}.

The large observed value of the top quark mass implies that
the product of all of the suppression factors 
on the \rhs of \Eq{mtbtc} must be
$\sim\! \frac{1}{10}$.
This is plausible only if
\emph{both} $y_t$ and $y_{\rm TC}$ are strong ($\sim\! 4\pi$) at the 
SUSY breaking scale.
We therefore require a theory where the technicolor gauge 
coupling, $y_{\rm TC}$, and $y_t$ are all strong at the
scale \MS.

This is not a coincidence if all of these couplings are
at strong conformal fixed points \emph{above} \MS.
Strong fixed-point Yukawa couplings occur naturally in 
SUSY conformal theories \cite{SUSYYuk}.
SUSY breaking is then a soft breaking of
conformal invariance that triggers the
transition from a supersymmetric conformal fixed point to a
non-supersymmetric conformal fixed point.

In order for $y_t$ to be at a strong supersymmetric fixed point,
the top quark must feel strong gauge interactions
above \MS.
This can naturally occur in models with an extended color
gauge symmetry.
We consider two models with color gauge symmetry breaking patterns:
\beq
\underbrace{SU(3)}_{\rm strong}
\times \underbrace{SU(3)}_{\rm weak}
\to\underbrace{SU(3)}_{\rm weak},
\eeq
and
\beq
\underbrace{SU(6)}_{\rm strong}
\times \underbrace{SU(3) \times SU(3)}_{\rm weak}
\to\underbrace{SU(3) \times SU(3)}_{\rm weak}.
\eeq
In the first case, the existence of a strong fixed point
does not allow all 3 generations of quarks to be charged under the
strong $SU(3)$.
We construct a model where the third generation is charged under
the strong $SU(3)$ (as in ``topcolor'' \cite{topcolor}).
In the second model, all three quark generations are charged
under the strong $SU(6)$.
Both models have additional flavor-dependent couplings,
but we show that they do not give unacceptably large flavor-changing
neutral currents for SUSY breaking scales as low as $\sim 10\TeV$.

The overall structure of our models is quite simple.
There are only two scales in the model, \MS and \LC,
and both have a natural origin.
Above the scale \MS the theory is supersymmetric,
and the technifermions and third generation quarks
(and possibly other quarks and leptons)
are at a strongly-interacting conformal fixed point.
SUSY breaks at the scale \MS, and below this scale all
the quarks and leptons become weakly interacting, while
the technicolor sector flows to a strongly-interacting
non-supersymmetric fixed point.
This conformal invariance in the technicolor sector is
softly broken by fermion mass terms at the scale \LC,
at which scale the technicolor dynamics confines
and breaks chiral symmetry.

One may expect that because $\MS \gg \mbox{TeV}$
there are no testable consequences of 
this framework at the TeV scale.
Remarkably, there is a robust testable consequence of
embedding conformal technicolor in a SUSY theory with
a strong conformal fixed point.
The strong conformal dynamics above \MS
suppresses the technicolor gaugino mass
\cite{softSUSYCFT}, so there is naturally a
$SU(2)_{\rm CTC}$ adjoint field below the TeV scale.
This gives rise to an additional pseudo Nambu-Goldstone
boson that can be observed at the LHC.

This paper is organized as follows.
In Section 2, we discuss the SUSY completion of the 
technicolor sector.
In Section 3, we discuss a SUSY model in which the strong
dynamics in the top sector arises from strong dynamics
in the third generation (``topcolor'').
In Section 4, we discuss a SUSY model in which 
all three generations are treated equally.
In Section 5, we discuss the phenomenology at the TeV
scale, particularly the PNGBs.
Section 6 contains some remarks on cosmology, and
Section 7 contains our conclusions.

% =============================================================================
\section{Supersymmetric Conformal Technicolor
\label{sec:SUSYCTC}}
% =============================================================================
In this section, we describe the technicolor sector above the SUSY
breaking scale \MS,
and how this matches onto minimal conformal technicolor
below \MS.
This sector is the same for both models of the strong
top dynamics that we discuss below.

% -----------------------------------------------------------------------------
\subsection{Field Content
\label{sec:SUSYCTCfields}}
The gauge group is
\beq
SU(3)_{\rm SCTC} 
\times SU(2)_L \times SU(2)_R
\eeq
with $Y = T_{3R}$.
The fields in the technicolor sector are
\beq[stechnifields]
\bal
\Psi &\sim (3, 2, 1),
\\
\tilde{\Psi} &\sim (\bar{3}, 1, 2),
\\
\Si_a &\sim (3, 1, 1),
\\
\tilde{\Si}_a &\sim (\bar{3}, 1 , 1),
\eal
\eeq 
with $a = 1, \ldots, 4$.
$SU(3)_{\rm SCTC}$ has 6 flavors,
and therefore has a maximally strongly-coupled (self-dual)
conformal fixed point \cite{SeibergSQCD}.

The technicolor gauge group has been extended from $SU(2)$
to $SU(3)$ for two reasons.
The first has to do with the anomalous dimensions of mass-squared
terms for technicolored scalars.
Mass-squared terms proportional to anomaly-free flavor symmetry
generators are not renormalized, while all other mass-squared
terms are suppressed \cite{softSUSYCFT}.
Therefore, some of the technicolored scalars 
have negative mass-squared at the scale \MS,
and hence partial breaking of the technicolor gauge symmetry is
inevitable.
Actually, because the technicolor gauge group is at a strongly
coupled fixed point, we cannot rigorously conclude that the
gauge group is spontaneously broken.
Here and throughout this paper we make the dynamical assumption
that strongly-coupled broken SUSY theories such as this are in
the same universality class as a weakly-coupled theory with the
same gauge group and matter content.
The second reason the gauge group is extended 
is that the partial breaking of the strong gauge group
fixes a mismatch between the number of
colors and flavors required for a strong fixed point.
For a strong SUSY fixed point, we want $N_c = 2N_f$,
while for a non-SUSY fixed point we probably want
$N_c \simeq 4 N_f$.
The latter estimate is very uncertain.
It is suggested by a number of lattice studies for
$N_c = 3$ \cite{CFTlattice12} 
(but see also \Refs{CFTdisagree}),
and agrees with model estimates \cite{CFTmodel}.
We assume VEVs of the form
\beq[SiVEV]
\avg{\Si}, \avg{\tilde{\Si}}
= \pmatrix{0 & 0 & 0 & 0 \cr
0 & 0 & 0 & 0 \cr
0 & 0 & 0 & * \cr}
\eeq
that break $SU(3)_{\rm SCTC} \to SU(2)_{\rm CTC}$
at the \MS.
Below \MS
$SU(2)_{\rm CTC}$ has 7 light fermion flavors (as we will explain below),
roughly where we expect a strong conformal fixed point.

There are an odd number of $SU(2)_{\rm W}$ doublets in \Eq{stechnifields},
so there is a global gauge anomaly.
We cancel this by adding the ``partner'' fields
\beq
\bal
P &\sim (1, 2, 1),
\\
\tilde{P} &\sim (1, 1, 2).
\eal
\eeq
The fields $\tilde{P}$ do not cancel any gauge anomaly, but they 
give rise to Dirac fermion masses, as we will see.

% -----------------------------------------------------------------------------
\subsection{Couplings to Flavor Messengers}
The higher dimension operators that give masses to quarks and leptons
arise from the exchange of heavy electroweak doublet scalars.
These have the quantum numbers of the MSSM Higgs fields,
namely
\beq[MSSMHiggs]
\Phi \sim (1, 2, 2).
\eeq
These fields have positive mass-squared,
and therefore do not get VEVs at the scale \MS.
We will refer to these fields as ``flavor messengers.''
Writing out the $SU(2)_R$ doublets explicitly, the
relevant fields have $SU(3)_{\rm SCTC} \times SU(2)_{\rm W} \times U(1)_{\rm Y}$
quantum numbers
\beq
\bal
\Psi &\sim (3, 2)_0,
\\
\tilde{\Psi}_1 &\sim (\bar{3}, 1)_{\frac 12},
\\
\tilde{\Psi}_2 &\sim (\bar{3}, 1)_{-\frac 12},
\\
\Phi_1 &\sim (1, 2)_{\frac 12},
\\
\Phi_2 &\sim (1, 2)_{-\frac 12}.
\eal\eeq
The superpotential couplings are
\beq[flavormesscoup]
\De W = y_{\rm{TC}1} \Psi \tilde{\Psi}_2 \Phi_1
+  y_{\rm{TC}2} \Psi \tilde{\Psi}_1 \Phi_2.
\eeq
At the strong fixed point, the operator $\Psi\tilde{\Psi}$ has dimension
$\sfrac 32$, so these couplings are relevant.
$\Phi_1$ couples to the top quark.
As discussed in the introduction, explaining the large value of the
top quark mass requires that $y_{\rm{TC}1}$ must be
strong at the scale \MS.
Below, we will 
demonstrate that this is not necessarily a coincidence.
In section \ref{sec:strongYukHid}, we will show that a mechanism similar to the Giudice-Masiero
mechanism for the $\mu$ term \cite{GM} can naturally generate relevant
interactions of this kind that get strong at \MS.
Additionally, in section \ref{sec:strongYukFix}, we will show that it is very plausible that relevant
Yukawa couplings, such as in \Eq{flavormesscoup}, run to a fixed point
when they become strong.

% -----------------------------------------------------------------------------
\subsection{Heavy Fermion Masses
\label{sec:HeavyFermionsSCTC}}
SUSY breaking automatically gives masses to all scalars,
but fermion masses must arise from superpotential couplings.
The ``3rd technicolor'' components of the fermions in
$\Psi$ and $\tilde{\Psi}$ get masses with the ``partner'' fields
$P$ and $\tilde{P}$ via the superpotential interactions
\beq[tc3masscoup]
\De W = y_P \tilde{\Si} \Psi P
+ y_{\tilde{P}} \Si \tilde{\Psi} \tilde{P}.
\eeq
At the strong fixed point, the operators
$\tilde{\Si} \Psi$ and $\Si\tilde{\Psi}$
have dimension $\frac 32$, so the couplings $y_P, y_{\tilde P}$
have dimension $\frac 12$.
These particles carry electroweak quantum numbers, so their
mass must be larger than of order $100\GeV$.
This means that we must require
\beq
\frac{100\GeV}{\MS^{1/2}} \lsim y_P, y_{\tilde{P}}
\lsim \MS^{1/2}.
\eeq
The same mechanism that explains why the couplings in 
\Eq{flavormesscoup} are strong near the scale \MS\ can work
for the couplings $y_P, y_{\tilde{P}}$ as well,
leading to masses of these fields near \MS.

The $SU(3)_{\rm SCTC}$ gaugino has a Majorana mass term from
SUSY breaking, but the Majorana mass operator $\la\la$ has 
scaling dimension $d> 3$ from the strong conformal dynamics
\cite{softSUSYCFT}.
The simplest SUSY breaking scenarios have SUSY broken at an energy
scale much larger than \MS, with SUSY breaking communicated to the
hidden sector by weak interactions.
In this case, the Majorana mass term will be 
strongly suppressed compared to other SUSY breaking terms
at \MS.
The VEVs, \Eq{SiVEV}, that break $SU(3)_{\rm SCTC}$ are of order
\MS, and give a mass to the $SU(3)_{\rm SCTC}$ gauginos
corresponding to broken gauge generators due to the super-Higgs
mechanism.
There is one linear combination of the $SU(2)_{\rm CTC}$ singlet
fermions in $\Si_4, \tilde{\Si}_4$ that is left massless.
(This can be understood from a counting argument: there are $8-3=5$
broken generators, but there are 6 fermions in $\Si_4, \tilde{\Si}_4$.)
This fermion can get mass from superpotential interactions
of the form
\beq
\De W \sim (\Si^{\vphantom{c}} \tilde{\Si})^2.
\eeq
This operator has dimension 4 at the fixed point.
This means that it can naturally be unsuppressed at the scale
\MS and give rise to masses of order \MS.

We also need small mass terms for sterile fermions to softly
break conformal symmetry at the TeV scale.
These arise from the superpotential couplings
\beq[Wsoft]
\De W \sim \Si\Si\Si + \tilde{\Si} \tilde{\Si} \tilde{\Si}
\eeq
where the $SU(3)_{\rm SCTC}$ indices are contracted with an
epsilon symbol.
Similarly, we may also require fermion masses for the
technifermions in $\Psi$ and $\tilde{\Psi}$ to control the
vacuum alignment below the electroweak breaking scale \cite{CTCmin}.
Such masses can arise from
\beq[Wsoft2]
\De W \sim \Si\Psi\Psi + \tilde{\Si} \tilde{\Psi} \tilde{\Psi}.
\eeq
These terms have dimension $\frac 94$, and are therefore also
relevant.
The coefficients of the superpotential terms \Eqs{Wsoft}
and \eq{Wsoft2}
must be small at the scale \MS, which is perfectly
natural.

% -----------------------------------------------------------------------------
\subsection{Strong Yukawa Couplings from Hidden Sector SUSY breaking
\label{sec:strongYukHid}}
We now discuss whether it is a coincidence for the relevant couplings
in \Eqs{flavormesscoup} and \eq{tc3masscoup} to be at or near
their strong coupling values at \MS.
This may be viewed as analogous to the $\mu$ problem in the
MSSM, which also requires a relevant supersymmetric interaction
to be important at the scale of SUSY breaking.
In this subsection, we present a simple model of hidden sector
SUSY breaking that naturally explains this coincidence using a
mechanism similar to the Giudice-Masiero mechanism for the $\mu$ problem.

We assume that SUSY is broken in the hidden sector by a chiral
superfield $X$ with $F_X \ne 0$.
We assume the usual couplings between $X$ and the visible sector
fields
\beq
\De K \sim \frac{1}{\MP^2} X^\dagger X q^\dagger q + \cdots
\eeq
where \MP is the Planck scale.
These give rise to visible sector SUSY breaking at the scale
\beq
\MS \sim \frac{\avg{F_X}}{\MP}.
\eeq
Suppose that in addition the hidden sector has a field $Y$ with
\beq[Ycond]
\avg{Y} \sim \avg{F_X}^{1/2},
\qquad
\avg{F_Y} \lsim \avg{F_X}^{1/2} \MS.
\eeq
Since $\avg{F_X}$ corresponds to the scale of SUSY breaking
dynamics, this is very natural as we will see below.
A coupling between the hidden and visible sectors of the form
\beq
\De W \sim \frac{1}{\MP^{1/2}} Y \Psi \tilde{\Psi}_2 \Phi_1
\eeq
then generates an effective Yukawa coupling
\beq
y_{\rm{TC}1} \sim \frac{\avg{Y}}{\MP^{1/2}} \sim \MS^{1/2}.
\eeq
This is just what we require in order for the Yukawa coupling
to be strong at the scale \MS.
The second condition in \Eq{Ycond} ensures that
visible sector SUSY breaking from $\avg{F_Y}$ is no
larger than \MS.

A simple hidden sector that accomplishes this has a superpotential
\beq
W = \ka X + \frac{\la}{\MP} Y^4.
\eeq
Note that this preserves a $Z_4$ symmetry under which
$Y \mapsto iY$.
We are looking for VEVs of the form \Eq{Ycond} with
$\avg{F_X} \sim \ka$, so we can expand around
$X, Y = 0$ in inverse powers of $\MP$.
In this case, a theorem due to Weinberg \cite{WeinbergSUGRA}
guarantees that if there is a vacuum with the desired properties
in the global SUSY limit, turning on SUGRA corrections will only
perturb the VEVs.
We therefore analyze the potential neglecting SUGRA corrections.
The most general $Z_4$ invariant \Kahler potential is therefore
\beq
\bal
K &= X^\dagger X + Y^\dagger Y
\\
&\qquad
+ \frac{1}{4 \MP^2} \left[
c_{XX} (X^\dagger X)^2
+ 4 c_{XY} X^\dagger X Y^\dagger Y
+ c_{YY} (Y^\dagger Y)^2 \right]
\\
&\qquad
+ \scr{O}(\MP^{-3}).
\eal
\eeq
where we have omitted holomorphic terms.
The potential is then
\beq
V = \frac{1}{\MP^2} \left[
-c_{XX} \ka^2 |X|^2 
-c_{XY} \ka^2 |Y|^2
+ \la |Y|^4 \right]
+ \scr{O}(\MP^{-3}).
\eeq
Assuming
\beq
c_{XX} < 0,
\qquad
c_{XY} > 0,
\eeq
the potential is minimized for
\beq
\avg{X} = 0,
\qquad
|\avg{Y}|^2 = -\frac{c_{XY}\ka^2}{2\la},
\eeq
which implies
\beq
\avg{F_X} = \ka,
\qquad
\avg{F_Y} = \frac{\la Y^3}{\MP}
\sim \frac{\avg{F_X}^{3/2}}{\MP} \sim \avg{F_X}^{1/2}\MS
\eeq
for $\la, c_{XY} \sim 1$.
If we ignore SUGRA corrections, this theory has a massless
``$R$-axion'' from the spontaneous breaking of a $U(1)_R$
symmetry under which $R(X) = 2$, $R(Y) = \frac 12$.
When we include SUGRA corrections, a crucial ingredient is
the addition of a constant term in the superpotential to
cancel the cosmological constant.
This explicitly breaks $U(1)_R$ and gives rise to a mass for the
$R$-axion \cite{SUGRARaxion}.
Of course there can be other sources of explicit $U(1)_R$ breaking
that also contribute to the $R$-axion mass.

% -----------------------------------------------------------------------------
\subsection{Strong Yukawa Couplings from New Fixed Points
\label{sec:strongYukFix}}
The conformal fixed point is no longer exact when a relevant
coupling like $y_{\rm{TC}1}$ is nonzero.
The strength of such couplings grows in the IR,
and causes the theory to flow away from the original fixed point.
A natural possibility is that the theory flows to a new
fixed point where the coupling $y_{\rm{TC}1}$ is itself at a strong
fixed point value.

There is a highly nontrivial consistency check on this hypothesis
from ``$a$ maximization'' \cite{amax}.
Provided that the superconformal $U(1)_R$ symmetry can be identified
with an anomaly-free $U(1)_R$ symmetry acting on fundamental
fields, the quantity
\beq
a = \frac{3}{32} \tr\left( 3 R^3 - R \right)
\eeq
is maximized over all allowed values of the $R$ charges.
This allows one to find the dimensions of all chiral operators
for a claimed fixed point.
These dimensions must be physical, for example the dimension
of scalar operators must be $\ge 1$.

The complete theory has strong dynamics in both the technicolor and
the strong sector, and the strong Yukawa coupling for the top quark
means that these sectors are also strongly coupled to each other.
Nonetheless, we will discuss the possible new fixed point in the
technicolor sector to illustrate the main points in a simpler setting.
We consider the case where only the coupling $y_{\rm{TC}1}$
is strong.
Here, we must maximize $a$ subject to the condition
$R(\Psi \tilde{\Psi}_2 \Phi_1) = 2$.
This gives a new fixed point with dimensions
($d = \frac 32 R$ for chiral operators)
\beq
d(\Psi) = 0.78,
\quad
d(\tilde{\Psi}_1) = 0.73,
\quad
d(\tilde{\Psi}_2) = 0.84,
\quad
d(\Phi_1) = 1.37.
\eeq
This is to be compared with the fixed point for small $y_{\rm{TC}1}$,
in which $d(\Psi) = d(\tilde{\Psi}_{1,2}) = \frac 34$ and
$d(\Phi_1) = 1$.
We see that the dimensions of the strongly interacting fields
have shifted only slightly, while the field $\Phi_1$ has a large
anomalous dimension in the new fixed point.
All gauge-invariant operators have physical dimensions,
and no scalar field has a dimension near 1, %JAE Edit
which is consistent with a strongly interacting fixed point.

Because the field $\Phi_1$ has dimension larger than 1 at the
new fixed point, weakly coupled Yukawa interactions involving
$\Phi_1$ are \emph{irrelevant} interactions at the new
fixed point.
The implications of this will be considered later when we discuss
complete models including quarks.

% -----------------------------------------------------------------------------
\subsection{Low Energy Effective Theory
\label{sec:LeffbelowMSUSY}}
For clarity and completeness, we summarize the effective
theory below the scale \MS.
The gauge group is
\beq
SU(2)_{\rm CTC} \times SU(2)_{\rm W} \times U(1)_{\rm Y}
\eeq
where the $SU(2)_{\rm CTC}$ is assumed to be at a strong
fixed point.
Supersymmetry is completely broken, and only gauge fields
and fermion fields survive below \MS.
The fermion fields are
\beq[fermionfieldsLeffCTC]
\bal
\psi &\sim (2, 1)_0,
\\
\tilde{\psi}_1 &\sim (2, 1)_{\frac 12},
\\
\tilde{\psi}_2 &\sim (2, 1)_{-\frac 12},
\\
\chi_a &\sim (2, 1)_0,
\\
\la &\sim (3, 1)_0,
\eal\eeq
where $a = 1, \ldots, 6$.
The $\psi$ and $\tilde\psi$ fields are the fermion components
of the superfields $\Psi$ and $\tilde\Psi$;
the $\chi_a$ are the fermion components of
$\Si_{1,2,3}, \tilde{\Si}_{1,2,3}$ (the components without VEVs);
and $\la$ arises from the $SU(3)_{\rm SCTC}$ gaugino.
The theory has mass terms for the fields $\chi, \psi, \tilde\psi$
arising from
the superpotential terms in \Eqs{Wsoft} and \eq{Wsoft2}:
\beq[Kmass]
\De\scr{L}_{\rm mass} = -\chi^T K \chi
-\ka \psi\psi - \tilde\ka \tilde{\psi}_1 \tilde{\psi}_2,
\eeq
where $K$ is an antisymmetric $6\times 6$ matrix.
There is also a mass term $\la\la$ arising from the
gaugino mass for the $SU(3)_{\rm SCTC}$ gauge multiplet,
but this is highly suppressed by the strong conformal
dynamics, as discussed in Section \ref{sec:HeavyFermionsSCTC}.
We assume that $K > \ka, \tilde\ka$, so that the conformal
invariance is softly broken by the $K$ terms.  
In fact, the $K$ mass terms arise from an operator \Eq{Wsoft}
that is more relevant than the operator \Eq{Wsoft2} that generates
the $\ka, \tilde\ka$ terms, so this is natural.
At the scale \LC where the $K$ terms get strong, the theory is
assumed to confine and spontaneously break the approximate
$SU(4)$ chiral symmetry acting on $\psi, \tilde\psi$.
The low-energy effective theory below the scale \LC
is discussed in detail in \Ref{CTCmin} for a model without 
the gaugino field $\la$.
The implications
of the additional light $\la$ will be discussed
in Section \ref{sec:pheno} below.

% =============================================================================
\section{Supersymmetric Topcolor
\label{sec:stopcolor}}
% =============================================================================
We now turn to the top sector above the SUSY breaking scale.
In this section, we describe the top sector in a model where only the third 
generation has strong dynamics above \MS.
This is similar to ``topcolor'' models \cite{topcolor},
so we refer to these models as ``supersymmetric topcolor.''

% -----------------------------------------------------------------------------
\subsection{Field Content}
The gauge group is
\beq
SU(3)_{\rm tC} \times SU(3)_{{\rm C}'}
\times SU(2)_{\rm W} \times U(1)_{\rm Y}.
\eeq
The quark fields have quantum numbers
\beq
\bal
q_3 &\sim (3, 1, 2)_{\frac 16},
\\
\tilde{t} &\sim (\bar{3}, 1, 1)_{-\frac 23},
\\
\tilde{b} &\sim (\bar{3}, 1, 1)_{\frac 13},
\\
q_i &\sim (1, 3, 2)_{\frac 16},
\\
% u_i^c 
\tilde{u}_i &\sim (1, \bar{3}, 1)_{-\frac 23},
\\
% d_i^c
\tilde{d}_i &\sim (1, \bar{3}, 1)_{\frac 13},
\eal
\eeq
where $i = 1, 2$ runs over the first two generations.
The breaking of the strong color group 
requires Higgs fields
\beq
\bal
\De &\sim (3, \bar{3}, 1)_0,
\\
\tilde{\De} &\sim (\bar{3}, 3, 1)_0.
\eal
\eeq
Mixing between the third generation and the first two
is mediated by an additional pair of vector-like
quarks:
\beq
\bal
U &\sim (1, 3, 1)_{\frac 23},
\\
\tilde{U} &\sim (1, \bar{3}, 1)_{-\frac 23},
\\
D &\sim (1, 3, 1)_{-\frac 13},
\\
\tilde{D} &\sim (1, \bar{3}, 1)_{\frac 13},
\eal
\eeq
From the fields listed so far,
$SU(3)_{\rm tC}$ has 5 flavors, a theory that is
dual to a $SU(2)$ gauge theory with 5 flavors.
This dual theory is likely weakly coupled, so we
add another ``junk''
flavor:
\beq
\bal
J &\sim (3, 1, 1)_{-\frac 13},
\\
\tilde{J} &\sim (\bar{3}, 1, 1)_{\frac 13}.
\eal
\eeq
These fields have the hypercharges of a
vectorlike down quark to avoid fractional electric charges.

% -----------------------------------------------------------------------------
\subsection{Color Symmetry Breaking}

The mass-squared term for the scalars charged under
the strong $SU(3)_{\rm tC}$ group must have both positive
and negative eigenvalues, as explained in Section \ref{sec:SUSYCTCfields}.
We assume that the negative eigenvalues result in VEVs of order
\MS of the form
\beq[DeltaVEVformtopcolor]
\avg{\De}, \avg{\tilde\De} \propto 1_3,
\eeq
resulting in the gauge symmetry breaking pattern
\beq
SU(3)_{\rm tC} \times SU(3)_{{\rm C}'} \to SU(3)_{\rm C}.
\eeq

% -----------------------------------------------------------------------------
\subsection{Heavy Fermion Masses
\label{sec:heavyfermionstopcolor}}
As noted previously, scalar masses of order \MS are automatically
generated for all fields.
We assume that all scalar mass-squared terms are positive for all fields
except for $\De$ and $\tilde\De$.
These fields get VEVs, but the
excitations at the minimum of the potential all have positive
mass-squared by definition.
On the other hand, fermion masses can only arise from superpotential terms.
In this subsection we describe the superpotential terms required
to give all unwanted fermions masses.

The fermion components of the Higgs fields $\De, \tilde{\De}$
and the ``junk'' flavors $J, \tilde{J}$
can get mass from superpotential terms
\beq[DeltaJmasscoupling]
\De W \sim (\De \tilde\De)^2
+ (\De \tilde\De) (J \tilde J).
\eeq
These couplings are marginal at the strong fixed point,
and therefore naturally generate fermion masses of order \MS.
Alternatively, they
can get masses from coupling to a singlet field $S$ via
\beq[SDeltacoupling]
\De W \sim S (\De \tilde{\De} + J \tilde{J}).
\eeq
The coupling of these has dimension $\frac 12$, and 
can be naturally strong at \MS via the mechanisms discussed
in Sections \ref{sec:strongYukHid} and \ref{sec:strongYukFix}.
The singlet $S$ can naturally get a VEV of order \MS,
so the mass of the fermionic components of $\De$ is naturally
of order \MS.

There are also a number of weakly-coupled
vectorlike fields.
These can get masses from ordinary $\mu$ terms
\beq
\De W \sim \Phi_1 \Phi_2 + J \tilde{J}
+ U \tilde{U} + D \tilde{D}
\eeq
or from couplings to the singlet field
$S$ in \Eq{SDeltacoupling}.

% ----------------------------------------------------------------------------
\subsection{Flavor
\label{sec:Flavortopcolor}}
We now study the flavor structure of this model.
The superpotential interactions that generate the quark masses are 
\beq[topcolorYukawa]
\bal
\De W &= (y_u)_{ij} q_i \Phi_1 \tilde{u}_j % u^c_j
+ y_t q_3 \Phi_1 \tilde{t}
+ (z_u)_i q_i \Phi_1 \tilde{U}
+ z_t U \tilde{\De} \tilde{t}
+ \mu_U U \tilde{U}
\\
&\qquad + (y_d)_{ij} q_i \Phi_2 \tilde{d}_j % d^c_j
+ y_b q_3 \Phi_2 \tilde{b}
+ (z_d)_i q_i \Phi_2 \tilde{D}
+ z_b D \tilde{\De} \tilde{b}.
\eal
\eeq
The usual $R$ parity assignments can be extended to the
fields in this sector to forbid relevant and marginal
$B$ and $L$ violating interactions.

The mass terms $\mu_{U,D}$ are assumed to be of order \MS,
and can be either explicit mass terms of order \MS
(\eg\ from the Giudice-Masiero mechanism)
or the VEV of the singlet field $S$ in \Eq{SDeltacoupling}.
We integrate out these fields to obtain the effective
flavor-dependent couplings below the scale \MS.
We define
\beq
\De_t \equiv z_t \avg{\tilde{\De}},
\qquad
\De_b \equiv z_b \avg{\De},
\eeq
and assume $\De_{t,b} \ll \MS$.
We obtain
\beq
\scr{L}_{\rm eff} &= \left[ (y_u)_{ij} q_i \tilde{u}_j % u_j^c 
+ y_t q_3 \tilde{t} 
- \frac{(z_u)_i \De_t}{\mu_U} q_i \tilde{t}
\right] \scr{H}_2^\dagger
\\
&\qquad{}
+ \left[ (y_d)_{ij} q_i \tilde{d}_j % d_j^c
+ y_b q_3 \tilde{b} 
- \frac{(z_d)_i \De_b}{\mu_D} q_i \tilde{b} 
\right] \scr{H}_1^\dagger,
\eeq
where we have defined
\beq
\scr{H}_1 = \frac{y_{\rm TC1}}{\MS^{d-1}} \psi \tilde{\psi}_1,
\qquad
\scr{H}_2 = \frac{y_{\rm TC2}}{\MS^{d-1}} \psi \tilde{\psi}_2.
\eeq
We see that this contains mixing between the third generation
and the first two generations.

This theory has flavor violation beyond minimal flavor violation.
To see this, we use a language where the heavy fermions mix with 
the light fermions (rather than an effective field theory language
where the heavy fermions are integrated out and their effects are
parameterized by higher-dimension operators).
The fermion mass terms are
\beq
\bal
\De\scr{L} &=
(m_u)_{ij} u_i \tilde{u}_j
+ m_t t \tilde{t}
+ \mu_U U \tilde{U}
+ \De_t U \tilde{t}
+ (\de_u)_i u_i \tilde{U}
\\
&\qquad
+ (m_d)_{ij} d_i \tilde{d}_j
+ m_b b \tilde{b}
+ \mu_D D \tilde{D}
+ \De_b D \tilde{b}
+ (\de_d)_i d_i \tilde{D} \, .
\eal
\eeq
where
\beq[mumddefn]
(m_u)_{ij} &= (y_u)_{ij} \avg{\scr{H}_2},
\qquad
m_t = y_t \avg{\scr{H}_2},
\qquad
(\de_u)_i = (z_u)_i \avg{\scr{H}_2},
\\
(m_d)_{ij} &= (y_d)_{ij} \avg{\scr{H}_1},
\qquad
m_b = y_b \avg{\scr{H}_1},
\qquad
(\de_d)_i = (z_d)_i \avg{\scr{H}_1}.
\eeq
We naturally have
\beq
m_{u,d}, \de_{u,d} \ll \De_{u,d}, \mu_{U,D}
\eeq
since the terms on the left are proportional to the electroweak symmetry
breaking scale, while those on the right are proportional to \MS.
In this language, the new flavor-violating effects come from
mixing with the ``fourth generation'' fields $U, \tilde{U}, D, \tilde{D}$.
Non-minimal flavor violation for the light generations is therefore
suppressed by the fact that the flavor violation arises from mixing through
a heavy fourth generation messenger.

We first discuss the constraints from $\De F = 2$ processes.
We focus on the down sector, which is the most sensitive.
The largest effect in this model comes from exchange of the 
massive color octet gauge bosons that arise from
$SU(3)_{\rm tC} \times SU(3)_{{\rm C}'} \to SU(3)_{\rm C}$.
These gives flavor-violating effects because the 3rd
generation fields $d_i, \tilde{d}_i$ have different $SU(3)_{\rm tC}
\times SU(3)_{{\rm C}'}$ quantum numbers from $b, \tilde{b}$.
The heavy gauge eigenstates are approximately the $SU(3)_{\rm tC}$
gauge bosons because these are strongly coupled.
The effective Lagrangian below the scale \MS therefore
contains the flavor-violating terms
\beq
\De\scr{L}_{\rm eff} \sim \frac{1}{\avg{\De}^2}
\tr\left( J^\mu_{\rm tC} \right)^2
\eeq
where
\beq
\left( J^\mu_{\rm tC} \right)_A = \sum_{I,J = 1}^3 \left[
(V_d)^*_{3I} (V_d)_{3J} d'^\dagger_I \si^\mu T_A d'_J
- (V_{\tilde{d}})^*_{3I} (V_{\tilde{d}})_{3J}
\tilde{d}'^\dagger_I \si^\mu T_A \tilde{d}'_J \right],
\eeq
where $T_A$ are $SU(3)$ generators, and the primed fields are the mass
eigenstates:
\beq
d'_I = (V_d)_{IJ} d_J,
\qquad
\tilde{d}'_I = (V_{\tilde{d}})_{IJ} \tilde{d}_J,
\eeq
with $I, J = 1, \ldots, 4$.
The flavor-violating effects are suppressed by the scale \MS,
and require mixing into and out of the third generation.
This cannot be too strongly suppressed if we hope to reproduce the
observed third generation mixing.

To estimate the mixing angles, we have found mass parameters that give an
approximate fit to the observed masses and CKM matrix elements.
A sample point of our viable space is given in Table~\ref{table:params}.
The resulting CKM matrix is found to be
\beq\label{eq:CKMtC}
V_{\rm CKM} \simeq \pmatrix {
0.97 & 0.24 &  0.003 \cr
0.24 & 0.97 &  0.04 \cr
0.008 & 0.04 & 0.99 }.
\eeq
All values are within 10\% or less of their experimental values,
which is sufficient for the estimates we make here.

\begin{table}[tb]
\begin{center}
\begin{tabular}{| c | c || c | c |}
\hline
\multicolumn{4}{|c|}{\hspace{0.13cm} Up Sector \hspace{0.7cm} \vline
 \hspace{0.6cm} Down Sector} \\
\hline \hline
$(m_u)_{11}$ &  $4.6 \times 10^{-3}$  & $(m_d)_{11}$ & $8 \times 10^{-3}$ \\
$(m_u)_{12}$ &  $7 \times 10^{-2}$ & $(m_d)_{12}$ &  $3.8 \times 10^{-2}$ \\
$(m_u)_{21}$ &  $6.1 \times 10^{-2}$ & $(m_d)_{21}$ & $ 3.6 \times 10^{-2}$ \\
$(m_u)_{22}$ & $1.5 $ & $(m_d)_{22}$ &  $0.13 $ \\
$m_{t}$ & 160  & $m_{b}$ &  5.7  \\
$(\delta_u)_{1} $ & $4.4$ & $(\delta_d)_{1} $ & $1.1$ \\
$(\delta_u)_{2} $ & $3.2$ &  $(\delta_d)_{2} $ & $2.8$ \\
$\mu_U$ & $14.2 \times 10^3$  & $\mu_D $ &  $11.7 \times 10^3$\\
$\Delta_t$ & $ 9.5 \times 10^3$  & $\Delta_b $ & $1.5 \times 10^3$ \\
\hline 
\end{tabular}
\caption{Values for quark masses (in GeV) that give an approximate fit to the
CKM matrix.
The corresponding CKM matrix is given in Eq.~(\ref{eq:CKMtC}).}
\label{table:params}
\end{center}
\end{table}

We then find
\beq
(V_d)_{31} &\sim \frac{(\delta_d)_1 \Delta_b}{\mu_D m_b} \simeq 2 \times 10^{-2},
\\
(V_d)_{32} &\sim \frac{(\delta_d)_2 \Delta_b}{\mu_D m_b} \simeq 6 \times 10^{-2},
\\
(V_{\tilde d})_{31} &\sim \frac{(\delta_d)_1 \Delta_b}{\mu_D m_b} \frac{m_d}{m_b} \simeq 4 \times 10^{-4},
\\
(V_{\tilde d})_{32} &\sim \frac{(\delta_d)_2 \Delta_b}{\mu_D m_b} \frac{m_s}{m_b} \simeq 2 \times 10^{-3}.
\eeq
Note that right-handed mixing is suppressed relative to that for left-handed fields.  The reason for this is simply that the fields $\tilde d_i$ are $SU(2)_W$ singlets, and so mixing into the singlet fourth generation through their coupling to $\Phi$ first requires a mass insertion for the light field.  This is in contrast to the left-handed case, where  the mass insertion must be included for the external {\it heavy} field.   

The strongest bound from these interactions comes from the
$B_s$--$\bar{B}_s$ mass splitting,
specifically the effective operator
$(b^\dagger \si^\mu s)^2/\Lambda^2$,
which requires $\La \gsim 130\TeV$ \cite{Isidori}.
To accomplish this with the parameters given in
Table~\ref{table:params}, we need
$\langle \Delta \rangle \gtrsim 8\TeV$.

Another potentially large contribution comes from
tree-level $Z$ exchange.
As we will show, this gives very weak bounds because it
requires mixing into and out of the \emph{fourth}
generation.
The $Z$ coupling is
\beq
\scr{L}_{\rm int} = g Z_\mu J^\mu_Z
\eeq
with
\beq
J^\mu_Z = \sfrac 12 (\cos\th_W - \sin\th_W)
(V_d)^*_{4I} (V_d)_{4J} d'^\dagger_I \si^\mu d'_J
+ \mbox{diagonal terms},
\eeq
where we have used the unitarity of the $4 \times 4$ mixing
matrix $V_d$.
The flavor-changing effective Lagrangian is therefore
\beq
\bal
\De\scr{L}_{\rm eff} &= -\frac{g^2}{8 m_Z^2}
(\cos\th_W - \sin\th_W)^2
(V_d)^*_{4I} (V_d)_{4J} 
(V_d)^*_{4K} (V_d)_{4L}
\\
&\qquad\qquad\qquad
\times (d'^\dagger_I \si^\mu d'_J)
(d'^\dagger_K \si_\mu d'_L).
\eal
\eeq
The mixing angles to the fourth generation are
\beq
\bal
(V_d)_{4i} &\sim \frac{(\de_d)_i}{\mu_D} \sim  10^{-4},
\\
(V_d)_{43} &\sim \frac{m_b \De_b}{\mu_D^2}  \sim 5 \times 10^{-5}.
\eal
\eeq
The strongest bounds for these operators come  from the $K$ system.
In particular, the imaginary part of the operator
$(s^\dagger \si^\mu d)^2/ \Lambda^2$ 
must satisfy $\La \gsim 1.7 \times 10^4\GeV$.
In the current case we see that such a limit provides no real constraint,
as the relevant mixing angles give an amplitude proportional to
$(\delta_d/ \mu_D)^4 \sim 10^{-16}$.
We conclude that there is no strong bound from these effects.

We have found that the strongest flavor bound on this model is
$\avg{\De} \gsim 8\TeV$.
We would like to translate this to a bound on \MS.
The field $\De$ has a quartic interaction of order
$g_{\rm tC}^2 \sim 16\pi^2$, so we have
\beq
\avg{\De} \sim \frac{\MS}{4\pi},
\eeq
since $\MS$ gives the scale of the SUSY breaking
mass-squared terms. 
We conclude that the bound is roughly $\MS \gsim 100\TeV$.

We close this section by noting that there are flavor-dependent
quartic superpotential interactions
\beq[ETCWtopcolor]
\De W \sim 
(q_3 \tilde{t}) (q_3 \tilde{b})
\eeq
that are marginal.
If these are unsuppressed at the scale \MS they can lead
to dangerous flavor-changing effects.
However, these terms can be forbidden by symmetries that
also forbid the ``$\mu$ term'' $\Phi_1\Phi_2$.
We expect such symmetries to be approximate symmetries
of the fundamental theory to explain why the $\mu$ term is
of order \MS.
We therefore expect that these operators will be strongly
suppressed in the fundamental theory.
In any case, it is technically
natural for the terms \Eq{ETCWtopcolor} to be suppressed
since they are superpotential interactions.

% -----------------------------------------------------------------------------
\subsection{New Yukawa Fixed Points
\label{sec:newfixedpttopcolor}}
Now that we have specified the top sector, we revisit the question
of why the top quark Yukawa coupling is strong at the scale \MS.
The most natural explanation is that the top quark Yukawa coupling
is at a strongly-coupled fixed point.
Assuming that both $y_t$ and  $y_{\rm{TC}1}$ flow to fixed 
points,
$a$-maximization (discussed in Section \ref{sec:strongYukFix}) 
gives the dimensions of the chiral fields:
\beq[topcolorfixedpt]
\bal
d(\Psi) = 0.77,
\quad
d(\tilde{\Psi}_2) = 0.80,
&\quad
d(\tilde{\Psi}_1) = 0.74,
\quad
d(\Sigma) =0.74, \\
d(q_3) = 0.77,
\quad
d(\tilde{t}) = 0.80,
&\quad
d(\Phi_1) = 1.44,
\quad
d(\Phi_2) = 1, \\
d(\tilde{b}) = d(J) &= d(\De) = 0.74.
\eal
\eeq
Note that the dimensions of the strongly coupled fields are close to $\frac 34$,
their value in the fixed point with vanishing Yukawa couplings, while
$\Phi_1$ has a large anomalous dimension.
This is consistent with the assumption that the fixed point is
strongly coupled.

Since $d(\Phi_1) = 1.44$, up-type Yukawa couplings of the first two
generations are irrelevant while the down-type Yukawa couplings
are marginal.
This fixed point cannot persist up to arbitrarily high scales,
otherwise the charm quark mass is too small.
We denote the scale where $\Phi_1$ gets a large anomalous
dimension by $\La_1$.
Since the charm Yukawa is an ordinary dimensionless coupling
at scale above $\La_1$, we require $y_c(\La_1) \lsim 1$.
Scaling down to \MS then gives
\beq
y_c(\MS) \lsim \left( \frac{\MS}{\La_1} \right)^{0.42},
\eeq
and the charm mass is given by
\beq
m_c \sim m_t \frac{y_c(\MS)}{y_t(\MS)}.
\eeq
Assuming $y_t(\MS)\sim 4\pi$, we find the bound
\beq
\frac{\La_1}{\MS} \lsim \left(
\frac{m_t}{4\pi m_c} \right)^{1/0.42}
\sim 10^3\, .
\eeq
We see that the scale $\La_1$ cannot be too far from the SUSY breaking
scale, so this new fixed point does not completely eliminate the need
for an explanation why relevant couplings get strong near the SUSY
breaking scale
(\eg\ the hidden sector discussed in 
Section \ref{sec:strongYukHid}).
However, these fixed points mean that
the coincidence is only within a few orders of magnitude
in the scale.

If we assume that $y_{\rm TC2}$ also flows to a strong fixed point,
we find
\beq[topcolorfixedpt2]
\bal
d(\Psi) = 0.80,
\quad
d(\tilde{\Psi}_2) = 0.77,
&\quad
d(\tilde{\Psi}_1) = 0.83,
\quad
d(\Sigma) =0.72, \\
d(q_3) = 0.77,
\quad
d(\tilde{t}) = 0.80,
&\quad
d(\Phi_1) = 1.42,
\quad
d(\Phi_2) = 1.37, \\
d(\tilde{b}) = d(J) &= d(\De) =  0.73.
\eal
\eeq
Now the Yukawa couplings for the leptons
and the first two generations of up- and down-type
quarks are irrelevant with roughly the same dimension
($d(y) \simeq -0.4$).
It is an appealing feature of this model that the masses of the 
light fermions are suppressed because they arise from irrelevant couplings.
This mechanism for flavor in conformal field theories was explored
in \Ref{NelsonStrassler} and also in the context of warped extra
dimensions (dual to conformal field theories) in \Refs{RSflavor}.
As with the previous fixed point, this fixed point cannot
be valid up to arbitrarily high scales.

We might ask whether the second fixed point requires
$y_t$ and $y_{\rm TC1}$ to become strong at precisely the same scale.
For example, if $y_{\rm TC1}$ becomes strong at a higher scale,
we have the fixed point discussed in Section \ref{sec:strongYukFix}.
In that fixed point $d(q_3 \tilde{t} \Phi_1) = 2.87$,
so the top coupling is still (barely) relevant, and therefore
continues to grow in the IR.
When it gets strong, we reach the second fixed point discussed above.

Another issue that arises in these new fixed points is the size of
the ``$\mu$ term'' $\De W \sim \Phi_1 \Phi_2$ that gives rise to the
mass of the fermion components of $\Phi_{1,2}$.
As above, we assume that $\Phi_1$ gets a large anomalous
dimension at the scale $\La_1$,
while $\Phi_2$ gets a large anomalous dimension at the
scale $\La_2 < \La_1$.
The size of the fermion mass term at the SUSY breaking scale
is then
\beq
\mu \sim \frac{\MS^2}{\La_1} 
\left( \frac{\La_1}{\La_2} \right)^{0.56}
\left( \frac{\La_2}{\MS} \right)^{0.21}.
\eeq
We must have $\mu \gsim 100\GeV$ to avoid conflict
with experiment.
To get a feeling for the bounds, we consider the case
$\La_1 = \La_2$ for simplicity.
We then obtain
\beq
\frac{\La_{1,2}}{\MS} \lsim \left( \frac{\MS}{100\GeV} \right)^{1/0.79}
\sim 5 \times 10^3 
\left( \frac{\MS}{100\TeV} \right)^{1.26}.
\eeq
We see that these bounds are similar to the ones that come
from requiring that the charm quark mass be sufficiently large.

Finally, note
that the superpotential couplings \Eq{DeltaJmasscoupling}
that give masses to the fermion components of $\De, \tilde\De, J, \tilde{J}$
are no longer exactly marginal in the new fixed points discussed here.
However, they are only slightly relevant (\eg\ the coupling
has mass dimension $+0.08$ for the second fixed point),
so it is still natural for these couplings to generate fermion masses
of order \MS.

% =============================================================================
\section{Supersymmetric Extended Color
\label{sec:SUSYEC}}
% =============================================================================
We now describe a different model in which all three generations of SM quarks are
charged under the same color group above \MS.
These models involve introducing additional colors of quarks,
so we refer to them as ``extended color'' models. 

% -----------------------------------------------------------------------------
\subsection{Field content}

The gauge group of the color sector is
\beq
SU(6)_{\rm EC} \times SU(3)_{\rm C1} \times SU(3)_{\rm C2} \times SU(2)_{\rm W} \times U(1)_{\rm Y}
\eeq
where only $SU(6)_{\rm EC}$ is strong at the scale \MS.
The quark fields are contained in the fields
\beq[quarksECmodel]
\bal
\scr{Q}_i &\sim (6,1,1,2)_{\frac 16}, \\
\tilde{\scr{U}}_i &\sim (\bar{6},1,1,1)_{-\frac 23}, \\
\tilde{\scr{D}}_i &\sim (\bar{6},1,1,1)_{\frac 13}, 
\eal
\eeq
where $i=1,2,3$ runs over all three generations.
The breaking of the color group is accomplished by the
Higgs fields
\beq
\bal
\De_1 &\sim (6,\bar{3},1,1)_0, \\
\tilde{\De}_1 &\sim (\bar{6},3,1,1)_0, \\
\De_2 &\sim (6,1,\bar{3},1)_0, \\
\tilde{\De}_2 &\sim (\bar{6},1,3,1)_0.
\eal
\eeq
To give masses to the extra color components of the quarks in \Eq{quarksECmodel}
we introduce three generations of ``partner quarks''
charged under $SU(3)_{\rm C2}$:
\beq
\bal
\tilde{Q}_i &\sim (1,1,\bar{3},2)_{-\frac 16}, \\
U_i &\sim (1,1,3,1)_{\frac 23}, \\
D_i &\sim (1,1,3,1)_{-\frac 13}, 
\eal
\eeq
where $i=1,2,3$.

It is easy to check that with the above
field content added to three generations of leptons
in the MSSM, all gauge anomalies cancel.
The $SU(6)_{\rm EC}$ gauge group has $12$ flavors,
just what we need for a strong conformal fixed point.
The gauge groups $SU(3)_{\rm C1} \times SU(3)_{\rm C2}$
are assumed to be weakly coupled at the SUSY breaking scale.

% -----------------------------------------------------------------------------
\subsection{Color Symmetry Breaking}
As explained in Section \ref{sec:SUSYCTCfields},
the mass-squared terms for scalars charged under the strong $SU(6)_{\rm EC}$
must have both positive and negative eigenvalues.
We assume that this gives rise to VEVs of order \MS
for the Higgs fields,
\beq
\avg{\De_1}, \avg{\tilde{\De}_1} \propto  \lp\begin{tabular}{c}
$1_3$ \\ $0_3$
\end{tabular} \rp,
\qquad
\avg{\De_2}, \avg{\tilde{\De}_2} \propto \lp\begin{tabular}{c}
$0_3$ \\ $1_3$
\end{tabular} \rp.
\eeq
This breaks
\beq
SU(6)_{\rm EC}\times SU(3)_{\rm C1}\times SU(3)_{\rm C2}
\to  SU(3)_{\rm C} \times SU(3)_{\mathrm{C}'}.
\eeq
The unbroken gauge groups $SU(3)_{\rm C} \times SU(3)_{\mathrm{C}'}$
are weakly coupled below \MS.
The ordinary quarks carry $SU(3)_{\rm C}$, which is therefore identified
with color.

There are no light particles charged under the unbroken
$SU(3)_{\mathrm{C}'}$.
This gauge group confines at a scale below \MS, and the heavy
particles charged under $SU(3)_{\mathrm{C}'}$ are ``quirks''
\cite{quirks} with masses of order \MS.
The weak gauging of $SU(3)_{\mathrm{C}'}$ does not play any important
role in this model, and we can turn off this gauging and break the
global $SU(3)_{\mathrm{C}'}$ symmetry without any important consequences
for physics below \MS.

% -----------------------------------------------------------------------------
\subsection{Heavy Fermion Masses}
As discussed in Section \ref{sec:heavyfermionstopcolor},
SUSY breaking gives masses of order \MS to all scalars,
but we need superpotential interactions to give masses to
unwanted fermion fields.

The ``extra'' $SU(6)_{\rm EC}$
colors of the quark fields $\scr{Q}_i, \tilde{\scr{U}}_i, \tilde{\scr{D}}_i$
get masses from
\beq[zcoupSEC]
\De W = (z_Q)_{ij} \scr{Q}_i \tilde{\De}_2 \tilde{Q}_j
+ (z_u)_{ij} \tilde{\scr{U}}_i \De_2 U_j
+ (z_d)_{ij} \tilde{\scr{D}}_i \De_2 D_j.
\eeq
The VEVs of $\De_2, \tilde{\De}_2$ then give masses to these
particles proportional to \MS.
These couplings have dimension $+\frac 12$ at the strong fixed point,
so they can naturally become strong near \MS via the
mechanism described in Section \ref{sec:strongYukHid}.
These new couplings violate flavor, and their impact on
flavor physics will be discussed below.

The fermion components of the Higgs fields
$\De_1, \tilde{\De}_1, \De_2, \tilde{\De}_2$
can get masses from the superpotential terms
\beq[quarticWfermionmassEC]
\De W \sim (\De_1 \tilde{\De}_1)^2
+ (\De_2 \tilde{\De}_2)^2
+ (\De_1 \tilde{\De}_1) (\De_2 \tilde{\De}_2).
\eeq
These terms are marginal at the fixed point we are considering.
Alternatively, they can arise from
coupling to a singlet field $S$:
\beq[singletWfermionmassEC]
\De W \sim S \De_1 \tilde{\De}_1
+ S \De_2 \tilde{\De}_2.
\eeq
The coupling of these terms has dimension $+\frac 12$,
so they can also naturally become strong near the scale \MS.
Alternatively, these couplings can themselves be at a
strong fixed point, as discussed in Section \ref{sec:strongYukFix},
and discussed further below.

% -----------------------------------------------------------------------------
\subsection{Flavor}
The quark masses arise from the Yukawa interactions
\beq
\De W = (y_u)_{ij} \scr{Q}_i \Phi_1 \tilde{\scr{U}}_j
+ (y_d)_{ij} \scr{Q}_i \Phi_1 \tilde{\scr{D}}_j.
\eeq
In addition, the superpotential couplings \Eq{zcoupSEC}
mix the quarks with heavy fermions at the scale \MS.
This generally gives rise to new flavor violation,
which we discuss in this section.

The fields $\scr{Q}_i, \tilde{\scr{U}}_i, \tilde{\scr{D}}_i$ 
decompose as
\beq
\scr{Q}_i = q_i + Q_i,
\quad
\tilde{\scr{U}}_i = \tilde{u}_i + \tilde{U}_i,
\quad
\tilde{\scr{D}}_i = \tilde{d}_i + \tilde{D}_i,
\eeq
where $q_i, \tilde{u}_i, \tilde{d}_i$ are the ordinary quark fields
with $SU(6)_{\rm EC}$ indices $1,2,3$,
and $Q_i, \tilde{U}_i, \tilde{D}_i$
are heavy fields with $SU(6)_{\rm EC}$ indices $4,5,6$.
All of these fields are triplets under the unbroken $SU(3)_{\rm C}$.
We further write the $SU(2)_{\rm W}$ doublets as
\beq
q_i = \pmatrix{u_i \cr d_i \cr},
\quad
Q_i = \pmatrix{ T_i \cr B_i \cr},
\quad
\tilde{Q}_i = \pmatrix{ \tilde{T}_i \cr \tilde{B}_i \cr}.
\eeq
As in Section \ref{sec:Flavortopcolor}, we find it convenient to
use a language where we do not integrate out the heavy fermions.
The mass matrix for the colored fields is then
\beq
\bal
W_{\rm mass} = (\De_Q)_{ij} \left[
T_i \tilde{T}_j
+ B_i \tilde{B}_j \right]
+ (\De_u)_{ij} \tilde{U}_i U_j
+ (\De_d)_{ij} \tilde{D}_i D_j
\\
+ (m_u)_{ij} \left[ u_i \tilde{u}_j
+ T_i \tilde{U}_j \right]
+ (m_d)_{ij} \left[ d_i \tilde{d}_j
+ B_i \tilde{D}_j \right],
\eal
\eeq
where
\beq
(\De_Q)_{ij} = (z_Q)_{ij} \avg{\tilde{\De}_2},
\quad
(\De_u)_{ij} = (z_u)_{ij} \avg{\De_2},
\quad
(\De_d)_{ij} = (z_d)_{ij} \avg{\De_2}.
\eeq
and $m_{u,d}$ are given by \Eq{mumddefn}.

There are $SU(6)_{\rm EC}$ gauge bosons that give rise to
transitions between ordinary quarks and heavy fermions.
The mass terms $\De_{Q,u,d}$ for the heavy fermions
are not generally diagonal in the same basis as the Yukawa
couplings, so this leads to flavor violation.
The most sensitive effects come from box diagrams with two
$SU(6)_{\rm EC}$ gauge boson exchange, which give \eg
\beq[LeffECexchange]
\De\scr{L}_{\rm eff} \sim 16\pi^2
\frac{[(\De_Q)_{12}]^2}{\MS^4} (s^\dagger \si^\mu d)^2,
\eeq
where $\De_Q$ is evaluated in the mass basis.
We have used $g_6 \sim 4\pi$ and used \MS for the mass of the
heavy gauge bosons.
This requires a high degree of alignment.
Assuming the effective operator $(s^\dagger \si^\mu d)^2 / \La^2$
has an unsuppressed imaginary part, we require
$\La \gsim 10^4\TeV$ \cite{Isidori},
giving
\beq[De12bound]
\frac{(\De_Q)_{12}}{\MS} \lsim \frac{\MS}{10^5\TeV}.
\eeq
For example, for $\MS \sim 100\TeV$ we require alignment $\sim 10^{-3}$.
This bound may be somewhat weakened if the new physics phase is aligned
with the standard model phase.
If we only use the bound for the $CP$-conserving part of the operator,
the bound is weaker by an order of magnitude.

We now argue that this high degree of alignment can arise very naturally.
Suppose that at some high scale there is a (gauged or approximate global)
flavor symmetry
\beq[flavorsymmetryEC]
SU(3)_Q \times SU(3)_{\tilde{U}} \times SU(3)_{\tilde{D}},
\eeq
with 
\beq
\bal
\scr{Q} &\sim (3, 1, 1),
\\
\tilde\scr{U} &\sim (1, \bar{3}, 1),
\\
\tilde\scr{D} &\sim (1, 1, \bar{3}),
\\
\tilde{Q} &\sim (\bar{3}, 1, 1),
\\
U &\sim (1, 3, 1),
\\
D &\sim (1, 1, 3).
\eal\eeq
Then the couplings $z_{Q,u,d}$ (and hence the masses 
$\De_{Q,u,d}$) do not violate the flavor symmetry \Eq{flavorsymmetryEC} as
long as they are proportional to the identity matrix
in flavor space.
The ordinary Yukawa couplings violate the flavor symmetry;
they can be viewed as spurions transforming under the flavor
symmetry as
\beq[Yukawaspurion]
\bal
y_u \sim (\bar{3}, 3, 1),
\qquad
y_d \sim (\bar{3}, 1, 3).
\eal
\eeq
Even if these are the only spurions that break the flavor symmetries,
there can be contributions to the couplings $z_{Q,u,d}$ that are
not proportional to the identity matrix, \eg
\beq
z_Q \sim (\bar{3} \times 3, 1, 1)
\sim 1 + y_u y_u^\dagger + y_d y_d^\dagger + \cdots
\eeq
However, these contributions are
diagonal in the basis that diagonalizes the Yukawa
couplings.
This is precisely the condition for minimal flavor violation.
Constructing detailed models of flavor is beyond the scope
of the present work, but we believe
that this flavor structure is quite reasonable.

There are also flavor-dependent
quartic superpotential interactions of the form
\beq[ETCWtopcolor]
\De W \sim (\scr{Q}\, \tilde{\scr{U}}) (\scr{Q}\, \tilde{\scr{D}})
\eeq
that are marginal.
If the flavor symmetry \Eq{flavorsymmetryEC} is broken only
by the Yukawa coupling spuions \Eq{Yukawaspurion}
then these will also be diagonal in the basis where the
Yukawa interactions are diagonal, and therefore
pose no problems for flavor.
In any case, it is technically
natural for the terms \Eq{ETCWtopcolor} to be suppressed
since they are superpotential interactions.

% -----------------------------------------------------------------------------
\subsection{New Yukawa Fixed Points}
We now address the question of whether Yukawa couplings involving 
strongly coupled fields (such as the top quark) can run to strong
fixed point values above the scale \MS.
Assuming that such Yukawa couplings do run to a fixed point, we can
use $a$-maximization (discussed in Section~\ref{sec:strongYukFix})
to find the dimensions of the chiral fields in the theory.
Assuming that $y_t$ and $y_{\rm{TC}1}$ are strong, we have 
\beq
\!\!\!\!\!\!\!\!
\bal
d(\Psi) &= 0.76,
\quad
d(\tilde{\Psi}_2) = 0.78,
\quad
d(\tilde{\Psi}_1) = 0.74,
\quad
d(\Sigma_a) = d(\bar{\Sigma}_a) = 0.74,
\\
&\ 
d(\scr{Q}_3) = 0.76,
\quad
d(\tilde{\scr{U}}_3) = 0.78,
\quad
d(\Phi_1) = 1.45,
\quad
d(\Phi_2) = 1,
\\
& \qquad\qquad
d(\De) = d(\scr{Q}_{1,2}) = 
d(\tilde{\scr{D}}_{1,2,3})= d(\tilde\scr{U}_{1,2}) = 0.75.
\eal
\eeq
In this fixed point the lepton Yukawa couplings are marginal,
so there is no problem getting lepton masses.
The Yukawa couplings of the light up-type quarks are
just barely relevant ($d(y_{u,c}) = 0.06$), while the Yukawa
couplings of all three generations of down-type quarks are very
relevant ($d(y_{d,s,b}) \simeq 0.5$).
It is consistent for this fixed point to
exist all the way up to the Planck scale, but then it would be
a complete coincidence that the up-type and down-type quark masses
have similar sizes (within an order of magnitude or two).
Also, the dimension of the Higgsino mass term $\Phi_1 \Phi_2$
is $1.45$, so there is no reason for the Higgsino mass to be
near \MS.
It is more natural to assume that this fixed point is reached
at a scale $\La_1$ not too far above \MS.
If we assume that above $\La_1$ there is an ordinary Higgsino 
mass term $\Phi_1\Phi_2$ of order \MS, then there is a bound on
the scale $\La_1$ from requiring that the Higgsino be heavier
than 100~GeV, but this bound is very weak:
\beq
\frac{\La_1}{\MS} \lsim 
\left( \frac{\MS}{100\GeV} \right)^{1/0.45}
\simeq 10^6 \left( \frac{\MS}{100\TeV} \right)^{2.2}.
\eeq
Requiring that the quark masses arise naturally is 
probably more restrictive, but it is difficult to quantify.

We can also consider a fixed point where
$y_t$, $y_{\rm{TC}1}$ and $y_{\rm{TC}2}$ are all at strong
fixed points.
In this case we find
\beq
\!\!\!\!\!\!\!\!
\bal
d(\Psi) &= 0.80,
\quad
d(\tilde{\Psi}_2) = 0.76,
\quad
d(\tilde{\Psi}_1) = 0.84,
\quad
d(\Sigma_a) =d(\bar{\Sigma}_a) =0.73,
\\
& d(\scr{Q}_3) = 0.77,
\quad
d(\tilde{\scr{U}}_3) = 0.79,
\quad
d(\Phi_1) = 1.45,
\quad
d(\Phi_2) = 1.37,
\\
& \qquad\qquad
d(\De) = d(\scr{Q}_{1,2}) = 
d(\tilde{\scr{D}}_{1,2,3})= d(\tilde\scr{U}_{1,2}) = 0.75.
\eal
\eeq
Now both the up- and down-type Yukawa interactions are barely
relevant, but the lepton Yukawa couplings are irrelevant.
Therefore, this fixed point cannot persist up to arbitrarily
high scales.
As in Section~\ref{sec:newfixedpttopcolor} this gives
a bound on the scale $\La_2$ where this new fixed point
is reached:
\beq
\frac{\La_2}{\MS} \lsim 
\left( \frac{m_t}{4\pi m_\tau} \right)^{1/0.37}
\simeq 150.
\eeq

We conclude that neither of these fixed points can be valid
to arbitrarily high scales (although for the first it is a
matter of naturalness).
We still need an explanation of why relevant couplings get
strong near the SUSY breaking scale
(\eg\ the hidden sector discussed in 
Section \ref{sec:strongYukHid}).
But now there is much less of a coincidence to explain.

% =============================================================================
\section{PNGB Phenomenology
\label{sec:pheno}}
% =============================================================================

All but one of the fields in the SUSY extensions to MCTC discussed in
Sections  \ref{sec:SUSYCTC}--\ref{sec:SUSYEC} can get masses
of order \MS, and therefore do not affect the phenomenology at the
TeV scale.
The one important exception is
the $SU(2)_{\rm CTC}$ component of the 
technicolor gaugino, which we call $\la$.
(See Section~\ref{sec:LeffbelowMSUSY} for a discussion of the effective
theory below the SUSY breaking scale.)
The gaugino mass term $\la\la$ has a large positive 
anomalous dimension in the SUSY conformal theory above the scale $\MS$
\cite{softSUSYCFT}.
If SUSY is broken in a hidden sector at a scale far above \MS and
communicated to the visible sector by weak interactions,
the coefficient of $\la\la$ will be very suppressed compared to soft
SUSY breaking terms with canonical dimensions.
Below the scale \MS, the operator $\la\la$ matches onto an operator
in the non-SUSY conformal technicolor theory, and we do not know
the scaling dimension of this operator.
However, it is very plausible that the $\la\la$ operator is still
suppressed at the scale \LC where conformal symmetry is broken.
For example in gravity-mediated SUSY breaking,
SUSY is broken at the intermediate scale $(\MS \MP)^{1/2}$,
while flavor bounds and the top quark mass suggest
$\MS \sim 100 \TeV$.
In this case, the suppression of $\la\la$ above \MS occurs
over 7 orders of magnitude, while the scaling between \MS
and \LC is two orders of magnitude.
While we cannot claim that a light gaugino at the scale \LC is 
an absolute prediction of this framework, it appears to be rather generic
in this framework.
We stress that this is not just a consequence of the specific models
presented here, but is expected in any model where conformal technicolor
is completed by a strong SUSY conformal theory.
As we have already argued, this is strongly motivated by the top
quark mass.
In this section, we therefore explore the phenomenological consequences
if the operator $\la\la$ is suppressed at the scale \LC
where conformal symmetry is broken.

In this case, the strong dynamics at the scale \LC
have an additional, approximate, anomaly-free $U(1)_\la$ global
symmetry:
\beq
\la \mapsto e^{-i\al} \la,
\qquad
\psi \mapsto e^{-i\al}\psi,
\qquad
\tilde\psi \mapsto e^{-i\al} \tilde\psi.
\eeq
We expect that $U(1)_\la$ will be spontaneously broken by
a gaugino condensate
$\avg{\la\la} \ne 0$,
so there is an associated PNGB that we call $\eta$.
In addition, there is an approximate global $SU(4)$ chiral
symmetry acting on the fields
\beq
\Psi^a = \pmatrix{ \psi \cr \tilde\psi \cr},
\eeq
that is spontaneously broken to $Sp(4)$ by a condensate
$\avg{\Psi^a \Psi^b} = -\avg{\Psi^b \Psi^a}$.
The scale of electroweak symmetry breaking depends on a vacuum
alignment angle $\th$, defined by
\beq
\avg{\Psi^a \Psi^b} \propto \Phi^{ab} = 
\pmatrix{\cos\th \, \ep & \sin\th\, 1_2 \cr -\sin\th \, 1_2 & -\cos\th \, \ep},
\eeq
where
\beq
\ep = \pmatrix{0 & 1 \cr -1 & 0 \cr}.
\eeq
Electroweak symmetry is broken at a scale
\beq
v = f \sin\th = 246\GeV.
\eeq
An important point is that the theory has an unbroken $SU(2)$
custodial symmetry for all values of $\th$ \cite{CTCmin}.
For $v \ll f$ we have a composite Higgs model \cite{compHiggs}.
The scale $f$ is related to the scale of conformal symmetry breaking
via $\LC \sim 4\pi f$.

There are 2 physical PNGB fields in $SU(4)/Sp(4)$
that can be parameterized by
\beq
\xi = e^{i\Pi/f},
\qquad
\Pi = h X_h + A X_A
\eeq
where
\beq
X_h = -\frac{i}{2} \pmatrix{ 0 & \ep \cr
\ep & 0 \cr},
\qquad
X_A = \frac{1}{2} \pmatrix{ \cos\th\, 1_2 & -\sin\th\, \ep \cr
\sin\th\,\ep & -\cos\th\, 1_2 \cr}.
\eeq
The field $h$ is $CP$-even, while $A$ and $\eta$ are $CP$-odd.
For $v \ll f$, $h$ is a composite Higgs.
We denote the additional PNGB arising from the spontaneous
breaking of $U(1)_\la$ by $\eta$.
It transforms as
\beq
\eta \mapsto \eta + 2 \al f_\eta
\eeq
where $f_\eta \sim f$ is the decay constant of the PNGB.
The field $\eta$ is $CP$-odd, so $\eta$ and $A$ will mix in general.

The potential for the PNGBs arises from interactions that explicitly
break the $SU(4) \times U(1)_\la$ symmetry.
We follow \Ref{CTCmin} and
consider the case where the PNGB potential 
is dominated by top quark loops and technifermion mass terms.
Justification and more detailed discussion of the potential
can be found in that paper.
We focus on the $CP$-odd PNGBs, since the potential for
$h$ is unchanged from \Ref{CTCmin}.
The technifermion mass terms \Eq{Kmass}
break $U(1)_\la$, and therefore give mass to the $\eta$. 
Defining
\beq
\mathcal K = \pmatrix{ \kappa \, \epsilon & 0 \cr 0 & \tilde \kappa \, \epsilon},
\eeq
the leading term in the potential is 
\beq[Vmasseff]
\!\!\!\!\!\!\!\!\!\!
V_{\rm mass} &= \sfrac 14 \hat{C}_\ka \tr(
\scr{K} \xi \Phi \xi^T) e^{i\eta/f_\eta}
+ \hc
\\
&= \hat{C}_\ka \left[ (\ka - \tilde\ka) \cos\th
\left(\frac{A^2}{f^2} + \frac{\eta^2}{f_\eta^2} \right)
+ 
%2\sqrt{2}\, 
(\ka + \tilde{\ka}) \frac{A\eta}{f f_\eta} 
+ \cdots \right].
\eeq
Here we assume that $\ka, \tilde\ka$ are real, and
\beq
\hat{C}_\ka \sim \frac{\LC^d}{16\pi^2}.
\eeq

The potential due to top quark loops does not break
$U(1)_\la$, and is therefore the same as in \Ref{CTCmin}.
Minimizing the total effective potential with respect to $h$ gives
\beq[thetavacuum]
\cos\th = \frac{\hat{C}_\ka (\ka - \tilde{\ka})}{C_t},
\eeq
where $C_t$ is an effective Lagrangian coefficient given by
\beq
C_t \sim \frac{3 m_t^2 \LC^2}{16\pi^2 \sin^2\th}.
\eeq
Combining this with the estimate of the top quark mass,
we find
\beq
m_h^2 = c_t N_c m_t^2
\eeq
with $c_t \sim 1$.
The $A$-$\eta$ mass matrix is then
\beq[Aetamassmatrix]
\De V_{\rm eff} = \frac 12\frac{m_h^2}{\sin^2\th}
\pmatrix{A \cr \eta \cr}^T
\pmatrix{1 & r_\ka r_\eta \cos\th \cr
r_\ka r_\eta \cos\th & r_\eta^2 \cos^2\th \cr}
\pmatrix{A \cr \eta \cr},
\eeq
where
\beq
r_\ka = \frac{\ka + \tilde\ka}{\ka - \tilde\ka},
\qquad
r_\eta = \frac{f}{f_\eta}.
\eeq
The potential has a minimum at $A, \eta = 0$ provided
that $|r_\ka| < 1$.
The mass eigenstates are
\beq
\pmatrix{A_2 \cr A_1 \cr}
= \pmatrix{\cos\al & -\sin\al \cr
\sin \al & \cos\al \cr}
\pmatrix{A \cr \eta \cr},
\eeq
where
\beq
\tan\al = \sfrac 12 \left( -x + \sqrt{x^2 + 4} \right),
\qquad
x = \frac{1 - r_\eta^2 \cos^2\th}{r_\ka r_\eta \cos\th}.
\eeq
The mass eigenvalues are
\beq
\!\!\!\!\!\!\!\!\!\!\!
m^2_{A_2, A_1} = \frac{m_h^2}{2\sin^2\th} \left[
1 + r_\eta^2 \cos^2\th
\pm \sqrt{(1 + r_\eta^2 \cos^2\th)^2 - 4 (1 - r_\ka^2) r_\eta^2 \cos^2\th}
\right].
\eeq
Note that
\beq
m_{A_1}^2 \le (1 - r_\ka^2)\frac{m_h^2}{\sin^2\th},
\eeq
so there is always one mass eigenstate lighter that
$m_h / \sin\th$, the value of the $CP$-odd PNGB mass
in the model without the light $\la$.

Note that in the limit $\tilde\ka \to 0$ ($\ka \to 0$)
where $r_\ka \to \pm 1$
one linear combination of $A$ and $\eta$ is massless,
as can be easily
seen from the determinant of the mass matrix \Eq{Aetamassmatrix}.
This arises because the mass terms preserve an anomaly-free $U(1)$
symmetry in this limit, namely (for $\ka = 0$)
\beq
\la \mapsto e^{-i\al} \la,
\qquad
\psi \mapsto e^{-2i\al}\psi,
\qquad
\tilde\psi \mapsto \tilde\psi.
\eeq
This limit is therefore completely natural.
In this limit, the mass eigenvalues are
\beq
m^2_{A_2} &\simeq \frac{m_h^2}{\sin^2\th}
\left( 1 + r_\eta^2 \cos^2\th \right),
\\
m^2_{A_1} &\simeq \frac{m_h^2}{\sin^2\th}
\frac{r_\eta^2 \cos^2\th}{1 + r_\eta^2 \cos^2\th} (1 - r_\ka^2),
\eeq
where
\beq
1 - r_\ka^2 =
\begin{cases}
\displaystyle
-4\tilde\ka/\ka & if $|\tilde\ka| \ll |\ka|$, \cr
\displaystyle 
-4\ka / \tilde\ka & if $|\tilde\ka| \gg |\ka|$. \cr
\end{cases}
\eeq

We now consider the couplings of the $CP$-odd PNGBs.
The $A$ field has suppressed couplings to fermions due
to the $Sp(4)$ custodial symmetry.
Specifically, we have  \cite{CTCmin}
\beq
g_{A\bar{f}f} \sim \frac{m_f}v \sin\th
\times N_c \left( \frac{ m_t}{4\pi v} \right)^2  r_\ka
 \cos\th.
\eeq
The first factor is the ``going rate'' for a PNGB with
decay constant $f = v/\sin\th$, while the second factor is
typically $\sim 10^{-2}$.
On the other hand, the $\eta$ field has unsuppressed
couplings to fermions via
the operator that generates fermion masses:
\beq
\scr{L}_{\rm eff} &= m_f f \tilde{f} e^{i\eta/f_\eta} + \hc
\\
&= m_f \bar{f} f + \frac{m_f}{f_\eta} \bar{f} i\ga_5 f \eta
+ \cdots,
\eeq
where the $\eta$ dependence is dictated by the fact that the
mass operator breaks $U(1)_\la$ by 2 units.
Neglecting the suppressed coupling of $A$ to fermions
we obtain couplings
\beq
\bal
g_{A_2 \bar{f}f} &= -\frac{m_f}{v} r_\eta \sin\th \sin\al,
\\
g_{A_1 \bar{f}f} &= \frac{m_f}{v} r_\eta \sin\th \cos\al,
\eal
\eeq
It does not appear natural to have either
$\sin\al$ or $\cos\al$ very small, and small
$\sin\th$ is the fine-tuned standard model limit.
Therefore, these couplings are generally comparable to the corresponding
standard model Yukawa couplings.

The dominant production mechanism for single $CP$-odd PNGBs 
at a hadron collider is 
therefore gluon fusion with a top loop or associated production
with $\bar{t}t$.
The dominant decay is to the heaviest
kinematically allowed fermion pair, just like the $CP$-odd
Higgs fields in multi-Higgs models.
This is challenging at the LHC, associated production with
$\bar{t}t$ and a boosted $A_{1,2}$ may be possible,
since this is similar to the analogous Higgs search discussed
in \Ref{ttbarHsearch}.
This has the possibility to measure the coupling of the scalar
to the top quark, and can therefore distinguish between the
$A_{1,2}$ and the Higgs.
Another possibility is pair 
production of $CP$-odd PNGBs via heavy resonances associated with
the strong dynamics at the scale \LC.
This is similar to strong double Higgs production in composite
Higgs scenarios, which may require very high integrated
luminosity \cite{DoubleHiggsPheno}.
We intend to investigate these signals in future work.

% =============================================================================
\section{Cosmology
\label{sec:cosmo}}
% =============================================================================
In this section, we make some brief remarks about the cosmology
of these models.
As written, the models have serious cosmological problems if the
reheat temperature of the universe is larger than
$\LC \sim \mbox{few TeV}$.
The reason is the theory at the scale $4\pi f$ contains resonances
with half-integer charge, for example resonances with the quantum
numbers $\psi\chi$ or $\tilde\psi\chi$ (see \Eq{fermionfieldsLeffCTC}).
Such particles are necessarily absolutely stable, since there is no
final state that they can decay to.
These particles are strongly interacting at the scale \LC,
and therefore have have large annihilation cross sections.
If their present cosmic abundance is determined by thermal freeze-out,
their abundance is somewhat less than that of dark matter.
Nonetheless, it is almost certain that such particles
are ruled out by the many constraints
on charged stable matter.
For example, direct searches for the flux of fractionally charged
particles $X$ at the earth
give a limit $\Phi_X \lsim 10^{-14}~\mbox{cm}^{-2}\,\mbox{s}^{-1}$
\cite{MACRO},
many orders of magnitude less than the expected flux for dark matter
\beq
\Phi_X \sim 10^8~\mbox{cm}^{-2}\,\mbox{s}^{-1}
\left( \frac{\Om_X}{\Om_{\rm DM}} \right)
\left( \frac{m_X}{\mbox{TeV}} \right)^{-1}.
\eeq
See \Refs{chargedDM} for additional constraints.

We can change the hypercharge assignments to eliminate
half-integer charged particles at the scale \LC.
For example, we can assign the particles $\chi$ in
\Eq{fermionfieldsLeffCTC} hypercharge $Y = \pm \frac 12$.
In this case, the particles that previously had half-integer
charges are still stable if we add no additional interactions,
but higher-dimension
interactions can be added to allow them to decay.
However, we then find that the SUSY model discussed in Section \ref{sec:SUSYCTC}
has half-integer charges that cannot be eliminated by
changing the hypercharge assignments.
We suspect that SUSY completions can be found
without fractionally charged particles, but we will not pursue
this here.

A simple solution to these difficulties is to assume that the
universe has a reheat temperature low enough that the troublesome
particles discussed above are never in thermal equilibrium.
This means a temperature below the temperature where they would
normally freeze out, of order 100~GeV for particles at the scale \LC.
This preserves the successes of big-bang nucleosynthesis, the highest
temperature direct experimental test of cosmology.
A cosmology with a low reheat temperature does not allow standard
mechanisms for dark matter and baryogenesis,
but there are simple and plausible
mechanisms in the literature that can be used in
the present context \cite{lowRHcosmo}.
We leave a full investigation of the cosmology of these models
to future work.

% =============================================================================
\section{Conclusions}
% =============================================================================
We have constructed complete models of flavor that reduce to minimal
conformal technicolor at low energies.
The models are based on supersymmetry broken at a high scale
$\MS \gg \mbox{TeV}$, and generate the required higher-dimension
operators via heavy scalar exchange.
We have argued that such a theory requires strong (super)conformal
dynamics at the scale \MS in both the technicolor and the top
sector.
In such a model, the top quark mass is given by
\beq
m_t \sim 4\pi v \left( \frac{4\pi f}{\MS} \right)^{d-1}
\eeq
where $d$ is the dimension of the ``Higgs'' operator
in the theory above the scale $4\pi v$,
and $f \gg v$ for a composite Higgs model.
We argued that the models are safe from flavor-changing neutral currents
for $\MS \gsim 100\TeV$.
Saturating this bound, the observed value of the top quark mass can be
obtained with $d \simeq 1.9$ for $f = v$.
Vacuum misalignment ($f > v$) can relax these bounds even more.

The models have a number of desirable features.
Most importantly,
there are only two mass scales in the theory, the SUSY breaking
scale \MS and the scale $\LC \sim 4\pi f \ll \MS$ where conformal
symmetry is broken and chiral symmetry breaking takes place.
Both of these scales arise naturally, with no fine-tuning.
Nontrivial SUSY fixed points or generalizations of the 
Giudice-Masiero mechanism can explain why supersymmetric mass
terms are near the SUSY breaking scale, so that all
unwanted fermions naturally get a mass of order \MS.

Another interesting result of this work is that this class of 
models makes a robust prediction for physics below the TeV scale,
namely the existence of an additional $U(1)$ PNGB.
This has couplings similar to a pseudoscalar Higgs, and may be
observable using jet substructure techniques.

A less attractive feature of the specific models we present is that they
require additional nontrivial flavor structure.
This arises from the sector responsible for making the top quark
strongly interacting at the SUSY breaking scale.
For the ``topcolor models'' of Section~\ref{sec:stopcolor},
additional flavor-dependent couplings are required to mix third
generation quarks with the first two generation.
For the ``extended color'' models of Section~\ref{sec:SUSYEC}
additional flavor dependent couplings are required to give mass
to the extra quark colors.
The need for this additional flavor structure 
arises for technical reasons, and is not obviously
inherent in our approach.
Specifically we require that the strong color group
$SU(N_c)$ gauge group have $N_f = 2N_c$ flavors
in order to be at a strong (self-dual) conformal fixed point.
In the ``topcolor'' model of Section~\ref{sec:stopcolor} 
the strong color group is $SU(3)$, but this is not strong if
all 3 generations feel the strong color force.
In the ``extended color'' model of Section~\ref{sec:SUSYEC}
we extend the strong color group to $SU(6)$, but then we need
additional flavor-dependent couplings to eliminate the 
extra colors of quarks.
All this depends on a ``Higgs'' picture of the dynamics where we
identify the fundamental degrees of freedom with the composite
quarks that emerge below the SUSY breaking scale.
This picture allows us to construct specific models, but 
it is quite possible that there are strongly coupled conformal
theories with 3 generations of composite fermions that do not
require flavor-dependent interactions beyond the usual Yukawa
couplings.
Finding such models would be very interesting.

However, the details of the models should not distract from the fact that
this work gives the first concrete proposal for a realistic
and UV complete dynamics of flavor
in a theory of strong electroweak symmetry breaking.
We believe that this removes a significant barrier to taking this
class of theories seriously as a framework for new physics at the
TeV scale.

% =============================================================================
\section*{Acknowledgements}
% =============================================================================
We thank S. Chang for discussions.
MAL thanks the Aspen Center for Physics for hospitality during
this work.
This work was supported by DOE grant
DE-FG02-91-ER40674.

\end{document}